\documentclass[aps,pra,amsmath,amssymb,twocolumn,superscriptaddress,notitlepage]{revtex4-2}
\usepackage{graphicx}
\usepackage{color}
\usepackage{braket}
\definecolor{burntorange}{rgb}{0.8, 0.33, 0.0}\usepackage[colorlinks,citecolor=blue,linkcolor=red,urlcolor=blue]{hyperref}
\hypersetup{
  pdfkeywords={},
  pdftitle={Criticality-enhanced Electric Field Gradient Sensor with Single Trapped Ions},
  }

\makeatletter

\newcommand{\ha}{\hat{c}}
\newcommand{\hadag}{\hat{c}^\dagger}

\newsavebox{\@brx}
\newcommand{\llangle}[1][]{\savebox{\@brx}{\(\m@th{#1\langle}\)}%
  \mathopen{\copy\@brx\kern-0.5\wd\@brx\usebox{\@brx}}}
\newcommand{\rrangle}[1][]{\savebox{\@brx}{\(\m@th{#1\rangle}\)}%
  \mathclose{\copy\@brx\kern-0.5\wd\@brx\usebox{\@brx}}}
\makeatother

\begin{document}

\title{Criticality-enhanced Electric Field Gradient Sensor with Single Trapped Ions}

\author{Theodoros Ilias}
\author{Dayou Yang}
\author{Susana F. Huelga} 
\author{Martin B. Plenio}
\email{martin.plenio@uni-ulm.de}
\affiliation{
Institut f{\"u}r Theoretische Physik and IQST, Universit{\"a}t Ulm,
Albert-Einstein-Allee 11, D-89069 Ulm, Germany}


\begin{abstract}
\noindent \textbf{ABSTRACT} 

\noindent We propose and analyze a driven-dissipative quantum sensor that is continuously monitored close to a dissipative critical point. The sensor relies on the critical open Rabi model with the spin and phonon degrees of freedom of a single trapped ion to achieve criticality-enhanced sensitivity.  Effective continuous monitoring of the sensor is realized via a co-trapped ancilla ion that switches between dark and bright internal states conditioned on a `jump' of the phonon population which, remarkably, achieves nearly perfect phonon counting despite a low photon collection efficiency. By exploiting both dissipative criticality and efficient continuous readout, the sensor device achieves highly precise sensing of oscillating electric field gradients at a criticality-enhanced precision scaling beyond the standard quantum limit, which we demonstrate is robust to the experimental imperfections in real-world applications.
\end{abstract}

\maketitle

\noindent \textbf{INTRODUCTION} 

\noindent A closed quantum many-body system is highly sensitive to even small variations in the Hamiltonian parameters near quantum critical points (CPs). This sensitivity is quantified by the divergent susceptibilities of its ground state and directly leads to a divergent quantum Fisher information (QFI), a key metric that characterizes the ultimately achievable precision of a many-body sensor. This observation has motivated manifold criticality-enhanced sensing proposals ~\cite{PhysRevA.78.042105,PhysRevA.88.021801,PhysRevX.8.021022,PhysRevA.96.013817,PhysRevLett.124.120504,dicandia2021critical,PhysRevLett.123.173601,PhysRevLett.126.010502,Salado_Mej_a_2021,PRXQuantum.3.010354,PhysRevResearch.4.043061,Ding_2022,PhysRevLett.130.240803}, offering promising alternatives towards quantum-enhanced sensitivity beyond conventional schemes that rely on the direct preparation of entangled states~\cite{PhysRevA.46.R6797,PhysRevLett.79.3865,Giovannetti1330}.

However, typical quantum sensors are intrinsically open systems both because they need to be interrogated by measurements and because they are naturally coupled to environmental noise. Hence, a natural next step for criticality-enhanced sensing is to look at open critical system undergoing \emph{dissipative phase transitions}~\cite{Baumann_2010,Klinder_2015,PhysRevLett.113.020408,PhysRevLett.118.247402,PhysRevX.7.011016,PhysRevX.7.011012,Cai_2022}. In analogy to the case of a closed system, at a dissipative CP the steady-state susceptibilities of an open system exhibit universally divergent scaling behavior, enabling the potential of achieving nonclassical precision scaling. Moreover, in sharp contrast to closed systems, open critical systems continuously exchange radiation quanta with their environment, making it possible to sense unknown parameters through continuous measurement of the emission field. This approach naturally avoids the detrimental effect of critical slowing down in steady state preparation ~\cite{PhysRevA.78.042105,PhysRevA.88.021801,PhysRevX.8.021022,PhysRevA.96.013817,PhysRevLett.124.120504,dicandia2021critical,PhysRevLett.123.173601,PhysRevLett.126.010502,Salado_Mej_a_2021,PRXQuantum.3.010354,PhysRevResearch.4.043061,Ding_2022,PhysRevLett.130.240803}, which needs to be factored into the resource budget of the metrology scheme thereby leading to a considerable reduction in the achievable precision \cite{PhysRevLett.130.170801}. 

In the innovative framework of sensing via continuous measurement~\cite{PhysRevA.87.032115,PhysRevA.89.052110,PhysRevA.94.032103,PhysRevLett.112.170401,Catana_2015,PhysRevA.64.042105,PhysRevA.93.022103,Schmitt832,PhysRevA.101.032347,Albarelli2018restoringheisenberg,PhysRevLett.125.200505,PhysRevA.93.032123,Clark2022exploitingnonlinear,Godley_2023,PhysRevX.13.031012}, the ultimate precision limit of an open sensor is characterized by the global QFI of the joint sensor and its environment. In a previous work \cite{PRXQuantum.3.010354}, we have established a general scaling theory of the global QFI at dissipative CPs, by relating its scaling exponents to the critical exponents of the underlying CP, assuming perfect photon detection and the absence of environmental noise. It remains, however, an important open question whether such scaling is robust against imperfect photon detection and various sources of experimental noise. Obtaining a definitive and positive answer to this question is of utmost importance for the successful implementation of our sensing protocol in real-world applications---particularly in view of the vulnerability of highly entangled states and their nonclassical precision scaling in the presence of experimental noise sources. Moreover, a direct implementation of the sensing protocol of Ref.~\cite{PRXQuantum.3.010354} requires highly efficient detection of the emission field emanating from engineered critical open systems. This can typically be accomplished through the use of high finesse optical cavities, which is, however, challenging experimentally~\cite{PhysRevLett.79.3865}.
\begin{figure}[t!]
\centering{} \includegraphics[width=0.48\textwidth]{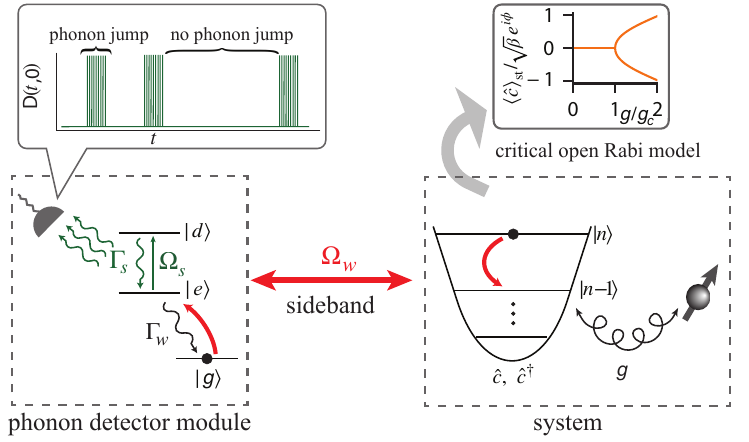} 
\caption{Schematic illustration of the proposed criticality-enhanced sensor. The sensor consists of a coupled bosonic mode and qubit system that realizes the critical open Rabi model. The quanta emitted into the bosonic mode can be monitored via a phonon detector module with high efficiency. In particular, via sideband coupling to an auxiliary spin, a single boson emission $\ket{n}\to\ket{n-1}$ switches the spin from the dark state $\ket{g}$ to the bright state $\ket{e}$ measured with standard electron shelving techniques. An additional decay channel $\ket{e}\to\ket{g}$ restores the internal state of the spin, allowing for continuous phonon counting at nearly perfect effective efficiency. Our measurement scheme can be viewed as an extension of the principle of quantum-logic spectroscopy for the implementation of continuous weak measurements. }
\label{fig:fig1} 
\end{figure} 

Here, we provide a confirmative answer to the first question and we overcome the second challenge with the proposal of a general cavity-free scheme for the highly efficient monitoring of driven-dissipative open quantum sensors, and demonstrate the persistence of criticality enhanced precision scaling for imperfect detection. Our scheme is inspired by entanglement-based spectroscopy \cite{doi:10.1126/science.1114375,Hemple, Wan} and is illustrated in Fig.~\ref{fig:fig1} at the example of detecting the emission quanta of a bosonic mode: to detect a single emission event we correlate the phonon states with an auxiliary spin-$\frac{1}{2}$ via sideband pulses that achieve the transition $\ket{n-1}\bra{n}\otimes\ket{e}\bra{g}$ ($\ket{n}$ is the bosonic Fock state) such that the event $\ket{n}\to\ket{n-1}$ is correlated with population transfer $\ket{g}\to\ket{e}$ of the auxiliary spin. The latter can be measured via standard electron shelving techniques with nearly unit efficiency \cite{RevModPhys.75.281}. The dissipative channel $\ket{e}\to\ket{g}$ of the spin restores its internal state thus enabling \emph{continuous} monitoring of the emission quanta of the bosonic mode with high efficiency. Natural extension of this scheme presents a versatile method for the continuous measurement of complex and interacting synthetic many-body quantum systems, which we demonstrate below using the critical open Rabi model realized with spin and phonons of trapped ions in a Paul trap. We will show that this system functions as a precise quantum sensor by harnessing driven-dissipative criticality and continuous monitoring, allowing for nonclassical precision scaling beyond the standard quantum limit. We showcase the criticality-enhanced precision scaling in terms of sensing the trapping frequency of the Paul trap, and demonstrate its robustness against realistic noise and imperfections. Since the trapping frequency directly reflects the electric field gradient at the ions' location, our proposed setup can serve as a promising precise sensor device suitable for electric-field imaging and diagnostics.

\vspace{7mm}

\noindent \textbf{RESULTS}

\noindent
\textbf{The Open Rabi Model as an Open Critical Sensor} 

\noindent We consider a bosonic mode coupled to a qubit according to the quantum Rabi Hamiltonian (throughout this article $\hbar=1$),
\begin{equation}
\label{eq:closed_Rabi}
\hat{H}_{\rm R}=\omega \hadag \ha +\frac{\Omega}{2} \hat{\sigma}_z-\lambda  (\ha+\hadag)\hat{\sigma}_x,
\end{equation}
where $\ha$ $(\hadag)$ denotes the annihilation (creation) operator of the mode, $\hat{\sigma}_{z,x}$ are the Pauli matrices of the qubit, $\omega$ is the mode's frequency, $\Omega$ is the qubit transition frequency and $\lambda$ is the coupling strength of the qubit-mode interaction. We assume that the bosonic mode is damped by a Markovian reservoir at a dissipation rate $\kappa$, and thus the dynamics of the open boson-qubit system can be described by a  Lindblad master equation (LME) \cite{PRXQuantum.3.010354,PhysRevA.97.013825,PhysRevLett.124.120504} 
\begin{equation}
\label{eq:LME_Rabi}
\dot{\rho}= -i [\hat{H}_{\rm R},\rho] 
+\kappa\left( 2 \hat{c} \rho \hat{c}^\dag 
-\{ \hat{c}^\dag\hat{c}, \rho \} \right).
\end{equation}
For such a zero-dimensional model, we can introduce the frequency ratio $\beta=\Omega/\omega$ as the effective system size, with $\beta\to \infty$ corresponding to the thermodynamic limit~\cite{PhysRevA.97.013825}. In the limit  $\beta \rightarrow \infty$, when the dimensionless coupling strength $g=2 \lambda /\sqrt{\omega \Omega}$ is tuned across the critical point (CP) $g_c=\sqrt{1+(\kappa/ \omega)^2}$, the steady state of Eq.~\eqref{eq:LME_Rabi} breaks spontaneously the ${\mathbb Z}_2$ parity symmetry ($\hat{c}\to-\hat{c},\hat{\sigma}_x\to-\hat{\sigma}_x$) of the model, therefore undergoing a continuous dissipative phase transition \cite{PhysRevA.97.013825}: from a normal phase for $g<g_c$, as characterized by the order parameter $\langle\hat{c}\rangle_{\rm st}=0$, with $\langle\cdot\rangle_{\rm st}\equiv{\rm tr}[\rho_{\rm st}(\cdot)]$; to a superradiant phase for $g>g_c$, where $\langle\hat{c}\rangle_{\rm st}\neq 0$. 

The CP is characterized by a few critical exponents which have been extracted via numerical finite-size scaling~\cite{PRXQuantum.3.010354,PhysRevA.97.013825}, in particular $z=1$ is the dynamic and $\nu=2$ is the correlation length critical exponent. Moreover, at the CP the mean excitation of the bosonic mode, $\hat{n}:=\hat{c}^\dag\hat{c}$, diverges as $\langle\hat{n}\rangle_{\rm st}\sim \beta^{1/2}$ which identifies the scaling dimension of $\hat{n}$ to be $\Delta_{\hat{n}}=-1/2$.  Consequently, in such model the standard quantum limit (SQL) corresponds to SQL $\sim \sqrt{\beta}$ while the Heisenberg limit (HL) is proportional to  HL $\sim \beta$. 

 In Ref.~\cite{PRXQuantum.3.010354} we had demonstrated the metrological potential of open critical systems subjected to continuous monitoring at the hand of an illustrative example of sensing the bosonic mode frequency, $\omega$, of the open Rabi model via photon counting. There, it was assumed that an annihilation of a phonon in the bosonic mode is associated with the emission of scattered photons all of which are directed to and counted by a photon detector with detection efficiency $\epsilon=1$. As a result, the dynamics of the joint boson-qubit system is subjected to measurement backaction conditioned on a series of photon detection events. Specifically, at any infinitesimal time step $d\tau$ there is probability $p_1=2\kappa \langle \hadag \ha \rangle_c d\tau$, with $\langle\cdots\rangle_c:=\langle\tilde{\psi}_c|\cdots|\tilde{\psi}_c\rangle/\langle\tilde{\psi}_c|\tilde{\psi}_c\rangle$, of photon detection, leading to the collapse of the conditional (unnormalized) state of the joint system, $|\tilde{\psi}_c\rangle\rightarrow \sqrt{\kappa d \tau}\ha |\tilde{\psi}_c\rangle$.  On the other hand, with probability $p_0=1-p_1$ we detect no photon and the system evolves according to the non-unitary evolution $|\tilde{\psi}_c\rangle\rightarrow \left(\hat{1}-d\tau(i \hat{H}_R+{\kappa} \hadag \ha) \right) |\tilde{\psi}_c\rangle$ \cite{RevModPhys.70.101, 1998aibp, carmichael1993open, wiseman_milburn_2009,gardiner2004quantum,gardiner2}. Repeating such a stochastic evolution defines a specific quantum trajectory, $D(t,0)$, consisting of the accumulated photon detection signal up to time $t$, with probability $P[D(t,0)]=\langle \tilde{\psi}_c(t)| \tilde{\psi}_c (t) \rangle$.

Processing the continuous signal $D(t,0)$ provides an estimator of the bosonic mode frequency whose achievable precision is represented by the Fisher information (FI) of the detected signal

\begin{equation} \label{eq:FI}
F_\omega(t) = \sum_{D(t,0)}P[D(t,0)]\left\{\partial_\omega {\rm{ln}} P[D(t,0)] \right\}^2.
\end{equation}
where the sum is over all, practically sufficient many, trajectories. Importantly, at the critical point $g=g_c$, the FI in Eq. \eqref{eq:FI} has been shown to exhibit a universal transient and long-time scaling behavior which surpasses the standard quantum limit \cite{PRXQuantum.3.010354}
\begin{align}
F_\omega(t,\beta) &= (\kappa t)^2 f_F(\kappa t/\beta), & \quad\kappa t\lesssim \beta, \label{eq:FI_Rabi_transient} \\
 F_\omega(t,\beta) &={\rm const.} \times\kappa t \beta, & \kappa t \gg \beta.\label{eq:FI_Rabi_long_time}  
\end{align}
where $f_F(\kappa t/\beta)$ is a universal scaling function reflecting the finite-size correction. 

The criticality-enhanced scaling, Eqs. \eqref{eq:FI_Rabi_transient} and  \eqref{eq:FI_Rabi_long_time}, demonstrates the potential of the critical open Rabi model as a quantum sensor when combined with highly efficient continuous measurements of the emitted quanta. Below we will describe a concrete realization of such a sensor model with trapped ions, where the model parameter $\omega$ translates directly to the actual trapping frequency of the ion, thus ultimately revealing the amplitudes and frequencies of unknown electric field gradients at ions' location. As mentioned in the introduction, a key achievement of our implementation scheme is the highly efficient detection of the emission quanta of the open sensor---corresponding to individual phonon excitation in our trapped-ion implementation---despite a low photon detection efficiency.

\begin{figure*}[t]
\centering{} \includegraphics[width=0.88\textwidth]{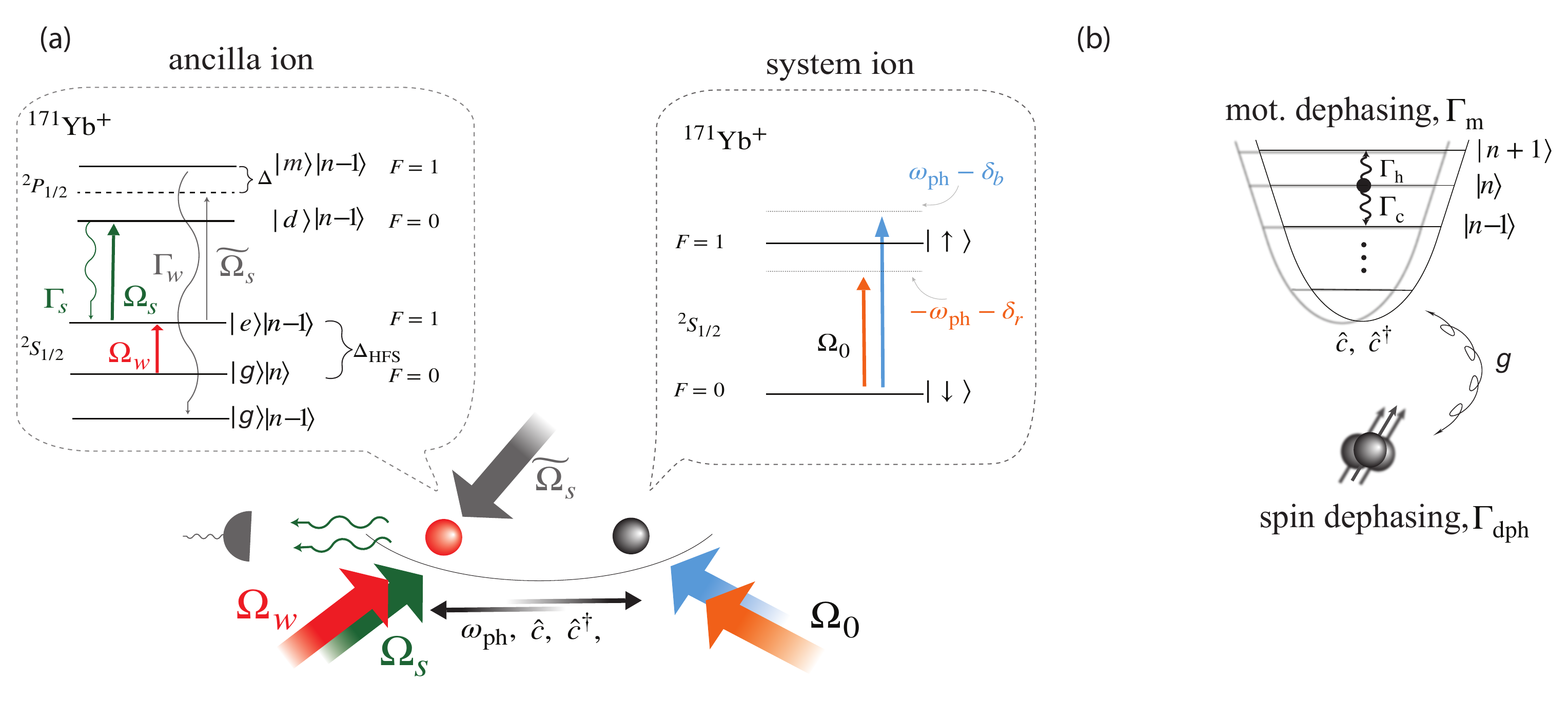} 
\caption {Implementation of the  critical quantum sensor using two co-trapped $^{171}{\rm Yb}^{+}$ ions in a linear Paul trap which share a common vibrational (phonon) mode. (a) The relevant energy levels are shown in the dashed boxes (not to scale). The system ion is driven by two laser beams slightly detuned by $\delta_{\rm b}$ and $\delta_{\rm r}$ from the blue and red sideband transition respectively, realizing the Rabi Hamitonian, $\hat{H}_R$. The ancilla ion is also driven by two laser beams of strength $\Omega_s$ and $\Omega_{w}$ which are tuned to the carrier resonance and slightly detuned from the red sideband transition respectively.  The annihilation of the phonon mode triggers the emission of multiple photons that are collected and counted by a photon detector. Finally, an additional laser of strength $\widetilde{\Omega}_s$, detuned by $\Delta$ from the auxiliary transition $|e\rangle \rightarrow |m \rangle$,  restores the internal state of the ancilla ion with a controllable rate and thus accomplishing continuous phonon detection with near unit efficiency. (b) The most detrimental noise sources in the trapped ion setup are:  i) motional diffusion which results in effective phonon heating and cooling with rate $\Gamma_{\rm h}$ and $\Gamma_{\rm c}$ respectively, ii) motional decoherence with rate $\Gamma_{\rm m}$ and iii) spin dephasing of the system ion at a rate $\Gamma_{\rm dph}$. }
\label{fig:fig2} 
\end{figure*}

\vspace{7mm}

\noindent
\textbf{Highly Efficient Continuous Monitoring of the Open Sensor}

\noindent A  promising platform for implementing our critical sensing scheme is a trapped-ion setup depicted in Fig.~\ref{fig:fig2}(a). We consider two trapped ions in a linear Paul trap sharing a quantized vibrational motion (phonon mode), which we assume is cooled down to the ground state. The two ions can be individually manipulated by lasers via single-ion addressability with focused laser beams  \cite{Linke3305} or via frequency space addressing with a crystal of mixed species \cite{Negnevitsky:2018ur}. As described in \cite{Pedernales:2015vl,PhysRevLett.118.073001,PhysRevX.8.021027,Cai:2021vx,PhysRevX.8.021027}, the Rabi Hamiltonian, $\hat{H}_{\rm R}$, can be implemented by driving the spin transition $\ket{\downarrow}\to\ket{\uparrow}$ of the system ion with two travelling-wave laser beams with the same Rabi frequency $\Omega_0$, and with frequencies slightly detuned from the blue and red-sideband transitions,

\begin{align}
\label{eq:laser_frequency}
\omega_1&=\omega_{\rm I}+\omega_{\rm ph}-\delta_{\rm b},\nonumber\\
\omega_2&=\omega_{ \rm I}-\omega_{\rm ph}-\delta_{\rm r}.
\end{align}
Here, $\omega_I$ is the frequency of the ionic spin transition, $\omega_{\rm ph}$ the center-of-mass (COM) phonon frequency and $\delta_{ \rm b (r)} \ll \omega_{\rm ph}$ is a small frequency offset. The transition $\ket{\downarrow}\to\ket{\uparrow}$ can represent a narrow optical transition subjected to laser driving as in $^{40}{\rm Ca}^{+}$; or a stimulated Raman transition between a pair of hyperfine ground states as in $^{171}{\rm Yb}^{+}$ \cite{Cai:2021vx}, which is illustrated in Fig.~\ref{fig:fig2}(a). After moving to a suitably rotating frame and performing an optical and a vibrational rotating-wave approximation (RWA), the Hamiltonian of the system ion reads (see Supplementary Note 2 for the detailed derivation)

\begin{equation}
\label{eq:ion_Rabi}
\hat{H}_{\rm R}=\frac{\delta_{\rm b}-\delta_{\rm r}}{2}\hat{c}^\dag\hat{c}+\frac{\delta_{\rm r}+\delta_{\rm b}}{4}\hat{\sigma}_z
-\frac{\eta_{\rm LD}\Omega_0}{2}\hat{\sigma}_x(\hat{c}+\hat{c}^\dag),
\end{equation}
where $\eta_{\rm LD} =k/\sqrt{2m\omega_{\rm ph}}$ is the Lamb-Dicke parameter, with $m$ the ion mass and $k$ the magnitude of the laser wavevector along the direction of the quantized oscillation. Eq.~\eqref{eq:ion_Rabi}  has exactly the same form as the Rabi Hamiltonian in Eq.~(\ref{eq:closed_Rabi}), with the new set of parameters $\omega=(\delta_{\rm b}-\delta_{\rm r})/2$, $\Omega=(\delta_{\rm b}+\delta_{\rm r})/2$ and $\lambda=\eta_{\rm LD}\Omega_0/2$. The last term of Eq. \eqref{eq:ion_Rabi} formally corresponds to a M\o lmer-S\o rensen interaction widely used for generating effective spin-spin interaction in an ion crystal~\cite{PhysRevLett.82.1971,Benhelm2008}. In contrast to Refs.~\cite{PhysRevLett.82.1971,Benhelm2008}, here we focus on individual trapped ions coupled with selected phonon modes. In our scenario, remarkably, Eq.~\eqref{eq:ion_Rabi} generates highly squeezed phonon states which ultimately manifests as a phase transition in the soft-phonon limit $\beta\equiv\Omega/\omega \rightarrow \infty$~\cite{PhysRevLett.115.180404}. Importantly, all the aforementioned parameters can be adjusted experimentally and thus allowing for tuning the spin-phonon system to the CP experimentally.

Hence, our  implementation realizes the parameter $\omega$ of the Rabi model as the differential detuning of the laser beams, which is directly related to the trapping frequency via Eq.~\eqref{eq:laser_frequency},
\begin{equation}
\omega=\frac{\delta_{\rm b}-\delta_{\rm r}}{2}\equiv\frac{\omega_2-\omega_1}{2} + \omega_{\rm ph}.
\end{equation}
As the frequency difference between the two lasers can be accurately controlled, estimating $\omega$ is equivalent to estimating the trapping frequency $\omega_{\rm ph}$. The access to $\omega_{\rm ph}$ provides us with a versatile tool for sensing unknown electric field gradients, as detailed in later sections.

Our sensing proposal further requires controlled dissipation of the phonon mode and efficient continuous detection of the phonons, both of which can be achieved via an ancilla ion, cf. Fig.~\ref{fig:fig1}.  In particular, we assume that the ancilla ion is driven by two additional laser beams: one slightly detuned from the red phonon sideband $| g\rangle |n\rangle \leftrightarrow | e\rangle | n-1\rangle$, with a detuning $\Delta_{w}=-\omega_{\rm ph}-\omega$; and a second one on resonance with the carrier transition $| e\rangle |n-1\rangle \leftrightarrow | d\rangle | n-1\rangle$, i.e., $\Delta_{s}=0$. The first laser drives the cooling transition with strength $\sqrt{n}  \eta_{\rm LD}^{(2)} \Omega_{\rm w}$, where $\eta_{\rm LD}^{(2)}$ is the Lamb-Dicke parameter of the ancilla ion; the second laser realizes the strong carrier transition with Rabi frequency $\Omega_{\rm s}$. In the regime $\Omega_{\rm s}, \Gamma_{\rm s} \gg \Omega_{ \rm w}, \Gamma_{\rm w}$, where $\Gamma_{\rm s (w)}$ is the  spontaneous emission rate of the strong (weak) transition $|d\rangle \rightarrow |e\rangle $ $( |e\rangle  \rightarrow |g\rangle )$, a controllable dissipation rate $\kappa \approx \Gamma_{\rm s} \Omega_{\rm w}^2/ \Omega_{\rm s}^2 $ is realized (see Supplementary Note 1), where each phonon annihilation is accompanied with the emission of a significant number of photons from the strongly driven transition which in turn are collected and counted by a photon detector with efficiency $\epsilon$.  We note that although in this section we assume that the state $|e\rangle$ relaxes directly to $|g\rangle$, in practice the weak $|e \rangle \rightarrow |g \rangle$ transition, necessary for the restoration of the internal state of the ancilla ion, may be accomplished through an auxiliary excited state $\ket{m}$, as illustrated in Fig.~\ref{fig:fig2}(a) and analyzed below. This results in the efficient continuous measurement of the phonon mode, featuring an enhancement factor (see Supplementary Note 3)
\begin{equation}
\label{eq:enhancement_factor}
N_{\rm ph}\equiv \frac{\text{detected photon \#} }{ \text{annihilated phonon \#} }\approx \epsilon \frac{\Gamma_{\rm s}}{\Gamma_{\rm w}}.
\end{equation}

Quantitatively, the conditional dynamics of the system and ancilla ions interacting with the phonon mode can be described by a stochastic master equation (SME) \cite{RevModPhys.70.101, 1998aibp, carmichael1993open, wiseman_milburn_2009,gardiner2004quantum,gardiner2} of the unnormalized joint state (see Supplementary Note 3)
\begin{align}
\label{eq:sme}
d\tilde{\rho}_{\rm c} =& \left(-i[\hat{H}_{\rm R}+\hat{H}_{\rm {a} },\tilde{\rho}_{\rm c}] \right. \\ \nonumber
& \left. -\frac{\Gamma_{\rm w}}{2}\{ |e\rangle \langle e|,\tilde{\rho}_{\rm c}\}+\Gamma_{\rm w}|g\rangle \langle e| \rho_{\rm c} ^{\prime} |e\rangle \langle g| \right.\\ \nonumber
&\left.-\frac{\Gamma_{\rm s}}{2}\{ |d\rangle \langle d|,\tilde{\rho}_{\rm c}\}+(1-\epsilon) \Gamma_{\rm s}|e\rangle \langle d|\rho_{\rm c} ^{\prime}  |d\rangle \langle e| \right) dt\\ \nonumber
&+ (\epsilon \Gamma_{\rm s} |e\rangle \langle d|\rho_{\rm c} ^{\prime}  |d\rangle \langle e|-\tilde{\rho}_{\rm c})dN(t),\nonumber
\end{align}
where $\hat{H}_{\rm a}=\Omega_{\rm s} (| d\rangle \langle e|+|e\rangle \langle d |)+\eta_{\rm LD}^{(2)}\Omega_{\rm w}(|e\rangle \langle g| \hat{c}+|g\rangle \langle e| \hat{c}^{\dag})$ is the Hamiltonian of the ancilla ion in the appropriate rotating-frame. Here, we have introduced the notation $\rho_{c\rm } ^{\prime}=\int_{-1}^1 du N(u) e^{-i \eta_{\rm LD}^{(2)}(\hat{c}+\hat{c}^{\dagger})u}\tilde{\rho}_{\rm c}e^{i \eta_{\rm LD}^{(2)}(\hat{c}+\hat{c}^{\dagger})u}$ with $N(u)$ being the angular distribution of the emitted photons. Such term accounts for the momentum transfer by spontaneous emission to the centre of mass motion of the ions when the excitation returns from the $|d \rangle$ to the $|e\rangle$ state or due to the transition $|e\rangle \rightarrow |g\rangle$. Although the effect of such process is explicitly analyzed later as one of the dominant noise sources of our detection scheme, in the remaining part of this section let us temporarily neglect it and study the ideal performance of our measurement scheme in the absence of noise. Moreover, we note that in Eq. (\ref{eq:sme})  $dN(t)$ is a stochastic Poisson increment which, similar to the ideal case of perfect detection, can take two values: if there is a photon detected, $dN(t)=1$ with a modified probability $\tilde{p}_1=\epsilon \Gamma_{\rm s}  {\rm tr} \{ |d\rangle \langle d|\rho_{\rm c}\}dt$, where $\rho_{\rm c}\equiv\tilde{\rho}_{\rm c}/{\rm tr}(\tilde{\rho}_{\rm c})$ is the normalized conditional density matrix, while if there is no photon detection $dN(t)=0$ with probability $\tilde{p}_0=1-\tilde{p}_1$. Taking the ensemble average over all conditional states and after elimination of the ancilla ion leads to the definition of the joint system ion--phonon state which evolves according to Eq. \eqref{eq:LME_Rabi}. 

\begin{figure}[t]
\centering{} \includegraphics[width=0.48\textwidth]{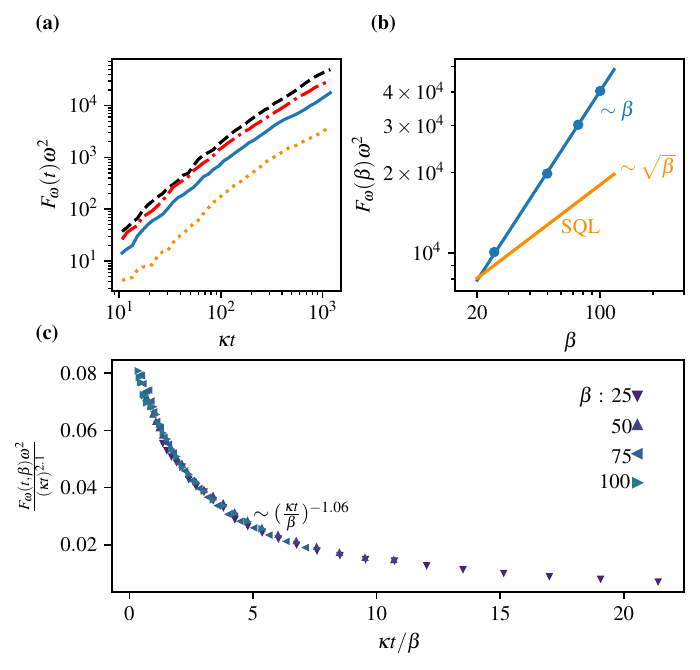} 
\caption{ Numerical estimation of Fisher information for photon counting signals with imperfect photon detectors. (a) The behavior of the FI, $F_{\omega} (t)$, in comparison with the ideal case of $\epsilon=1$ (black dashed line) for $\epsilon=0.1$, $\Gamma_{\rm s}/ \Gamma_{\rm w}=80 \rightarrow N_{\rm ph}\approx 8$ (red dashed-dotted line); $\epsilon=0.01$, $\Gamma_{\rm s}/ \Gamma_{\rm w}=80 \rightarrow N_{\rm ph}\approx 0.8$ (blue solid line);  $\epsilon=0.01$, $\Gamma_{\rm s}/ \Gamma_{\rm w}=10 \rightarrow N_{\rm ph}\approx 0.1$ (orange dotted  line). In all curves $\Gamma_{\rm w}=40$, $\Omega_{\rm w}=14.3\Gamma_{\rm w}$, $\Omega_{\rm s}=2\Gamma_{\rm s}$ and $\eta_{\rm LD}^{(2)}=0.07$. The above set of parameters for the ancilla ion gives rise to a phonon dissipation rate $\kappa=1.5 \times 10^{-3}\Gamma_{\rm w}$. The system ion is tuned to the dissipative CP and we choose $\omega=10 \kappa$, $\beta=50$. (b) Long time behavior of the achieved FI with respect to the effective system size $\beta$ (blue line). For comparison we plot the scaling of FI that corresponds to SQL \cite{PRXQuantum.3.010354} (orange line) obtained with the use of a coherent state of the bosonic model (see main text).  (c) Finite size scaling analysis of $F_{\omega} (t)$ averaged over $10^4$ trajectories for the case of $\epsilon=0.01$ and $\Gamma_{\rm s}/ \Gamma_{\rm w}=80$. The remaining parameters are the same as in Fig. ~\ref{fig:fig3}(a).}
\label{fig:fig3} 
\end{figure}

The enhanced metrological precision via our measurement scheme is demonstrated in Fig.~\ref{fig:fig3}(a), which shows the FI for the estimation of the frequency of the phonon mode, $F_{\omega} (t)$, at various enhancement factors, $N_{\rm ph}$.
As illustrated, increasing $N_{\rm ph}$ leads to larger FI and thus enhanced precision, which can in principle achieve the ideal precision for perfect photon detection. In the experimental relevant case of photon detector efficiency  $\epsilon=0.01$ and $N_{\rm ph} \approx 80 \epsilon$, we perform a numerical finite-size scaling analysis for different experimentally accessible system sizes $\beta$ (see below),  as shown in Fig.~\ref{fig:fig3}(c) where the perfect data collapse indicates that the $F_{\omega} (t)$ for $\epsilon_{\rm ph}<1$ follows a similar transient and long time behavior as in Eqs. \eqref{eq:FI_Rabi_transient} and \eqref{eq:FI_Rabi_long_time}. In the following we analyze  a concrete experimental realization of our measurement scheme, shown in Fig.~\ref{fig:fig1}, and we examine its robustness against various experimental noise sources.

\vspace{7mm}

\noindent
\textbf{Parameters for Experimental Implementation} 

\noindent Various ion species can be used for implementing our measurement scheme, and here, as a concrete example, we consider  $^{171}{\rm Yb}^{+}$ as the ancilla ion, whose internal level structure is shown in Fig.~\ref{fig:fig2}(a). We choose from the ground $^2 S_{1/2}$ manifold the hyperfine states $| g\rangle\equiv| { F}=0,{m_F}=0 \rangle$ and $| e \rangle\equiv| {F}=1,{m_F}=0 \rangle$ to implement the phonon sideband cooling, cf. Fig.~\ref{fig:fig1}, with a frequency difference $\Delta_{\rm HFS} \approx 2 \pi \times 12.6 {\rm GHz}$, while the readout state is chosen  as $|d \rangle\equiv| {F}=0,{m_F}=0 \rangle$ from the excited $^2 P_{1/2}$ manifold. The cooling transition $| g\rangle |n\rangle \rightarrow | e\rangle | n-1\rangle$ can be accomplished either by exploiting techniques utilizing microwave fields on resonance with the red phonon sideband \cite{Ospelkaus2011,PhysRevLett.87.257904}, or more straightforwardly by stimulated Raman transition at a two-photon detuning $\Delta_{\rm HFS}-\omega_{\rm ph}$ \cite{Blinov}.  The strong cycling transition $\ket{e} \longleftrightarrow$ $ \ket{d}$, with a spontaneous emission rate $\Gamma_{\rm s}=2 \pi \times 19.6 {\rm MHz}$  of the $^2P_{1/2} $ manifold, can be driven by a laser beam at $369.5 {\rm nm}$ with adjustable Rabi frequency $\Omega_{\rm s}$, cf. Fig.~\ref{fig:fig2}(a). Since the $\ket{d}\to\ket{g}$ emission channel is forbidden by dipole selection rules, an additional laser of strength $\widetilde{\Omega}_{\rm s}$, detuned by $\Delta$ from the $\ket{e} \to \ket{m}  \equiv$ $^2P_{1/2} | {F}=1,{ m_F}=0 \rangle$ transition, induces an effective decay $|e\rangle \rightarrow | g\rangle$ at an adjustable rate \cite{Noek:13},

\begin{equation}
\label{eq:weak_emission_rate}
\Gamma_{\rm w} \approx \frac{\Gamma_{\rm s}}{2} \frac{(2 \widetilde{\Omega}_{\rm s}/\Gamma_{\rm s})^2}{1+(\frac{2 \widetilde{\Omega}_{\rm s}}{\Gamma_{\rm s}})^2 +(\frac{2 \Delta}{\Gamma_{\rm s}})^2 }.
\end{equation}

As described in \cite{PhysRevX.8.021027,Cai:2021vx}, the two states of the ground state manifold of a second $^{171}{\rm Yb}^{+}$ ion can be chosen as the qubit states, which together with the spatial motion of the ion along one of its principle axis, with frequency $\omega_{{\rm ph}}=2 \pi \times 2.35 {\rm MHz}$, provide the two degrees of freedom for implementing the Rabi Hamiltonian. A typical set of experimental parameters is $\omega \approx 2 \pi \times 2 {\rm kHz}$ and $ (\eta_{\rm LD} \Omega_0)_{\rm max} \approx 2 \pi \times 20 {\rm kHz}$ with $\eta_{\rm LD}=0.07$. Therefore, effective system size of $\beta \lesssim 100$ can be easily achieved. To achieve even larger $\beta >100$, a stronger Rabi frequency $\Omega_0$ is required to tune the system to the CP, which may ultimately break the vibrational RWA and modifies the resulting Hamiltonian of the system ion. This can be overcome by suppressing the corresponding carrier transition, e.g, by using standing wave configuration~\cite{PhysRevLett.118.073001}  or exploiting the ac Stark shift of travelling waves~\cite{PhysRevA.62.042307} to implement the sideband transitions, which allows for exploring the critical physics close to the thermodynamic limit. As a possible implementation of the ancilla-enhanced continuous readout, we consider that $\eta_{\rm LD}^{(2)} \Omega_{\rm w} \approx 2 \pi \times 150 {\rm kHz}$ with $\eta_{\rm LD}^{(2)}=0.07$ and $\Gamma_{\rm w}$ can be adjusted to $\Gamma_{\rm w} =\Gamma_{\rm s}/80= 2 \pi \times 245 {\rm kHz}$. These realistic parameters result in a phonon dissipation rate $\kappa=2\pi \times 200 {\rm Hz}$ with $N_{\rm ph} \approx 80 \epsilon$ (see Supplementary Note 3) which, remarkably, allow for a clear demonstration of the criticality enhanced precision scaling as shown in Fig.~\ref{fig:fig3}(b).

\begin{figure}[h]
\centering{} \includegraphics[width=0.48\textwidth]{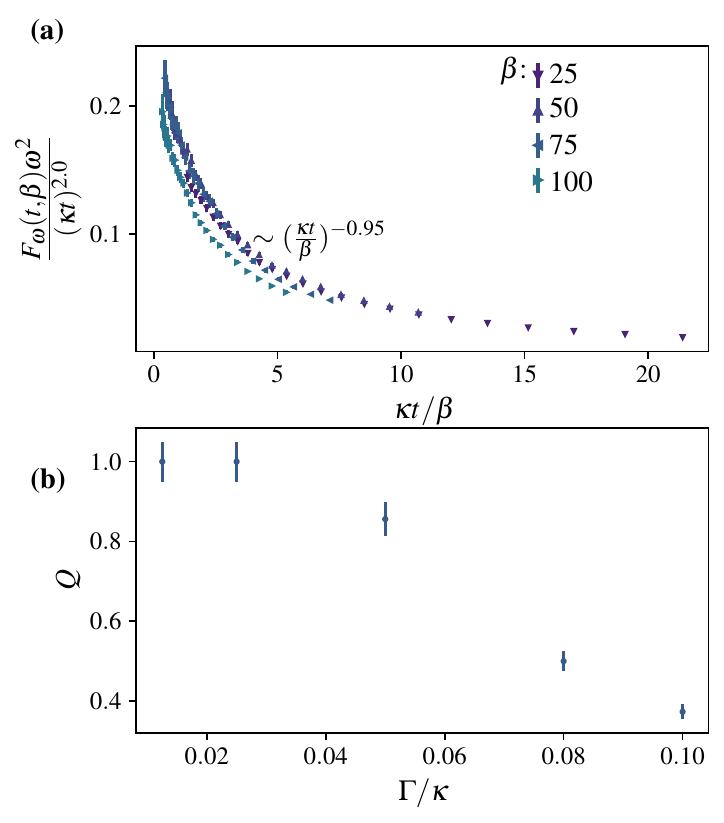} 
\caption{(a) Finite size scaling analysis of $F_\omega(t)$ for $\epsilon=1$ including noise with $\Gamma_{\rm dph}=\kappa/200$ and $\Gamma_{\rm m}= \Gamma_{\rm h}=\Gamma_{\rm c}=\kappa/10$. The rest of the parameters are the same as in Fig.~\ref{fig:fig3}. Each data point represents an average over $5 \times 10^3$ independent trajectories. (b) The quality of data collapse, $Q$, in comparison with the noiseless scenario. For $\Gamma \leq \kappa/40$, the Fisher information,  $F_\omega(t)$, follows the same scaling behaviour as predicted for the noiseless case.}
\label{fig:fig4} 
\end{figure}

\vspace{7mm}
\noindent\textbf{Robustness Against Experimental Noise}

\noindent Noise can be detrimental to any quantum enhanced sensing protocol and certainly our proposal would be incomplete without a systematic study of the impact of realistic noise sources, found in the experiments, in the performance of our proposed sensor setup. As illustrated in Fig. \ref{fig:fig2}(b), the main relevant experimental noise sources  are (i) spin decoherence of the system ion at an effective rate $\Gamma_{\rm dph}$, including also the effect of laser frequency noise, (ii) motional decoherence of the phonon mode at a rate $\Gamma_{\rm m}$, and (iii) motional diffusion of the phonon mode mainly caused by photon recoil in the strong transition $|d \rangle |n\rangle \rightarrow |e\rangle |n-1\rangle$ of the ancilla ion (at a rate $\Gamma_{\rm s}$, cf. Fig.~\ref{fig:fig2}). In the sideband resolved regime, such photon recoil results in effective phonon heating and cooling, $|d \rangle |n\rangle \rightarrow |e\rangle |n\pm1\rangle$, at a rate $\Gamma_{\rm h}$ and $\Gamma_{\rm c}$ respectively. The rates are given by $\Gamma_h\approx \Gamma_c \approx \frac{2}{5}R (\eta_{\rm LD}^{(2)})^2$ where $R=\frac{N_{\rm ph}}{\epsilon} \kappa$ is the rate of the scattered photons and $\frac{2}{5}\eta_{\rm LD}^{(2)}$ is the probability to decay in a sideband \cite{PhysRevA.46.2668}. Note that all the above imperfections do not affect either the effective dissipation rate of the phonon mode $\kappa$ or the highly efficient ancilla-assisted continuous phonon counting. Consequently, for the sake of numerical efficiency, we examine the effect of noise under the assumption of perfect photon detection efficiency $\epsilon=1$. In this case the SME of the unnormalised  system ion-phonon state reads
\begin{align}
\label{eq:SME_noise}
d\tilde{\rho}_{\rm c}=& \Big( -i[\hat{H}_{R},\tilde{\rho}_{\rm c}]+\Gamma_{\rm dph} \mathcal{L} [\hat{\sigma}_z]+\Gamma_{\rm m} \mathcal{L} [\hat{c}^{\dagger} \hat{c}] \nonumber \\
&+ \Gamma_{\rm h} \mathcal{L} [\hat{c}^{\dagger}]+\Gamma_{\rm c} \mathcal{L} [\hat{c}] -\kappa \{ \hat{c}^{\dagger} \hat{c}, \tilde{\rho}_{\rm c} \} \Big) dt \nonumber \\
&+(2 \kappa \hat{c} \tilde{\rho}_{\rm c} \hat{c}^{\dagger}-\tilde{\rho}_{\rm c} )dN(t) 
\end{align}
where $\mathcal{L}[\hat{x}]\equiv \hat{x} \tilde{\rho}_{\rm c} \hat{x}^{\dagger}-\frac{1}{2} \{ \hat{x}^{\dagger} \hat{x}, \tilde{\rho}_{\rm c}\}$. We take typical experimental numbers $\Gamma_{\rm dph}=2 \pi \times 1{\rm Hz}$ and $\Gamma_{\rm m}= \Gamma_{\rm h}=\Gamma_{\rm c}=\Gamma=2 \pi \times 20 {\rm Hz}$~\cite{Cai:2021vx,Islam:14}. Consequently, fixing $\Gamma_{\rm dph}=\kappa/200$ and $\Gamma=\kappa/10$ while keeping the rest of the parameters the same as Fig.~\ref{fig:fig3}, we show the resulting FI $F_\omega(t)$ in Fig.~\ref{fig:fig4}(a). The results indicate that although the various noise sources degrades the perfect data collapse, a scaling behavior similar to the noiseless case persists. The quality of the data collapse can be further quantified by the dimensionless quality factor $Q$~\cite{Somendra} that captures the mean relative spread among different sets of data (see Methods). We identify that for the current state-of-the-art experimental setups the main impact emanates from the motional decoherence and diffusion. As shown in Fig.~\ref{fig:fig4}(b), $Q\sim 1$ persists for small motional decoherence rate, indicating that the associated Fisher information follows the same scaling behavior as the noiseless case. This demonstrates the feasibility of our sensing scheme under realistic experimental imperfections.

\vspace{7mm}

\begin{figure}[h!]
\centering{} \includegraphics[width=0.48\textwidth]{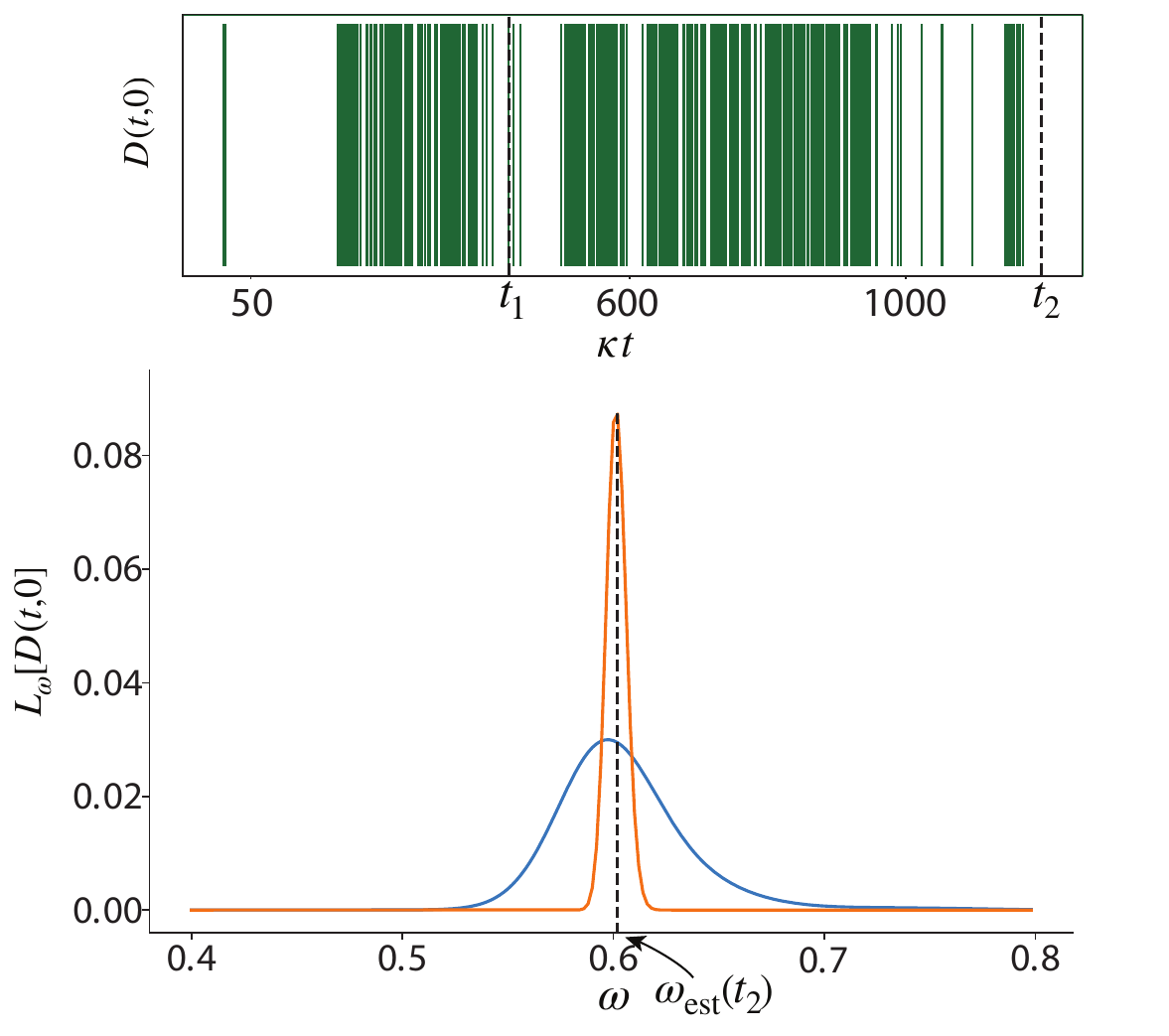} 
\caption{Extracting the information of the frequency of the bosonic mode via continuous measurement. We consider sensing the bosonic mode's frequency $\omega$ via the detection of the emitted photons from the strong transition of the ancilla's ion as described in the main text. Processing the simulated continous signal $D(t,0)$ (upper panel) via Eq. (\ref{eq:sme}) allow us to construct the likelihood function $L_{\omega}[D(t,0)]$ shown for $\{\kappa t_1, \kappa t_2 \}=\{424, 1197 \}$. In turn, a maximum-likelihood strategy provides us with an estimate value  $\omega_{\rm est}(t)$ shown here for $t=t_2$. The rest of the parameters are the same as in Fig. \ref{fig:fig3} with $\epsilon=0.01$.}
\label{fig:fig5} 
\end{figure} 

\noindent\textbf{Performance for Sensing External Electric Field Gradients}

\noindent The proposed sensor utilizes the center-of-mass  phonon mode of the co-trapped ions along the radial $x$-direction to implement the bosonic degrees of freedom of the Rabi Hamiltonian, Eq. (\ref{eq:ion_Rabi}), and the associated dissipation and continuous readout. In a linear Paul trap, the radial trapping frequency $\omega_{\rm ph}$ is determined via the trap parameters as
\cite{RevModPhys.75.281}
\begin{equation}
\label{eq:phonon_freq}
\omega_{\rm ph}=\sqrt{ \frac{4 q}{m \Omega_{\rm rf}^2} \left(\frac{U_r}{R^2}-\gamma U_0\right)+\frac{1}{2}\left(\frac{2qV_0}{\Omega_{\rm rf}^2mR^2}\right)^2}\frac{\Omega_{\rm rf}}{2}.
\end{equation}
Here, $q$ and $m$ is the charge and the mass of the ions, $R$ is the distance between the trap center and the surface of the electrodes and $\gamma$ is a geometric factor. As in standard linear Paul traps, we consider the application of a radio-frequency potential $V=U_r+V_0 \cos(\Omega_{\rm rf}t)$ between diagonally opposite electrodes of the trap, with  $U_r(V_0)$ being its dc(ac) component. This results in a quadrupolar potential near the trap center $\Phi_{\rm rf}=\frac{1}{2}\left[U_r+V_0 \cos(\Omega_{\rm rf}t)\right](1+\frac{x^2-y^2}{R^2})$. Meanwhile, a dc voltage $U_0$ in the outer segments of the electrodes realizes a static trapping potential $\Phi_{\rm st}=\gamma U_0  (z^2-\frac{x^2+y^2}{2})$. Consequently, our scheme is capable of sensing variations in the pseudopotential's curvature, $\omega_{\rm ph}$, resulted either by small drifts of the frequency $\Omega_{\rm rf}$ due to different technical noise sources, or external electric field gradients (static or oscillating at frequency $\Omega_{\rm rf}$) at the ions' location.

To verify the high sensitivity of the proposed sensor, let us analyze its performance under typical parameter choices as in Fig. \ref{fig:fig3} and an experimentally achievable effective system size $\beta\simeq 50$. From Fig. \ref{fig:fig3}(a), it is clear that for the experimentally reachable system size $\beta=50$ we obtain $F_{\omega} \omega^2 \approx 36 \times 10^2 $ for a realistic evolution time $\kappa t \approx 200$.  As a result, the sensitivity for measuring the model parameter $\omega$ is $\frac{\Delta \omega}{\omega} =\frac{1}{F_\omega \omega}\approx 1.6 \times 10^{-2}$. This directly translates to the  sensitivity of measurement of the trap frequency
 \begin{equation}
 \label{eq:sensitivity}
 \frac{\Delta \omega_{\rm ph}}{\omega_{\rm ph}}=\frac{\Delta \omega}{\omega}\frac{\omega}{\frac{\omega_1-\omega_2}{2}+\omega} \approx 1.6 \times 10^{-5},
 \end{equation}
 where $\omega_{1 (2)} \sim 2 \pi \times 12.6 {\rm GHz}$ for $^{171}{\rm Yb}^{+}$, and we have assumed an accurate control of the two laser beams in implementing the Rabi Hamiltonian. We note that such sensitivity is comparable to the one achieved using excited motional Fock states in a single run of experiments as described in Ref. \cite{Wolf2019}. In such approach the  trap's frequency, $\omega_{\rm z}$, can be estimated by detecting a residual displacement $\tilde{\alpha}$ where the interrogation time of each experimental run lasts $t_f\approx 7.8 {\rm ms}$. Consequently, using the reported experimental numbers and taking into account that in our main text we have assumed a scheme lasting $t\approx 0.16 {\rm s}$, leads to sensitivity $|\frac{\Delta \omega_{\rm z}}{\omega_{\rm z}}| \approx 6 \times 10^{-6}$ which is similar to the one achieved with our protocol as derived in Eq. (\ref{eq:sensitivity}). 

Furthermore, our scheme, besides its use for estimating the trap frequency and its direct connection with electric field gradients, can be particularIy useful for measuring other physical parameters of interest characterizing the potential applied in the electrodes. For example, if we are interested in estimating $U_0$ and assuming $U_r=0$, using Eq. (\ref{eq:phonon_freq}) we straightforwardly obtain 
 \begin{equation}
 |\frac{\Delta U_0}{U_0}|=8 \frac{\Delta \omega_{\rm ph}}{ \omega_{\rm ph}}| \frac{ \omega_{\rm ph}}{A-4 \omega_{\rm ph}^2 }|,
 \end{equation}
 where $A :=\frac{\Omega_{\rm rf}^2}{2}(\frac{2qV_0}{\Omega_{\rm rf}^2mR^2})^2=4(\omega_{\rm z}^2-\omega_{\rm ph}^2)$ with $\omega_{\rm z}=\sqrt{2q \gamma U_0/m}$ being the frequency of the center-of-mass mode in the axial direction. Using typical experimental numbers where $\omega_{\rm z}/\omega_{\rm ph} \approx 0.1$ \cite{Debnath2016} we obtain
\begin{equation}
|\frac{\Delta U_0}{U_0}| \approx 3.2 \times 10^{-3}.
\end{equation}

Importantly, in our proposal the Fisher information and thus the precision of estimating the unknown parameter $\omega$, or equivalently $\omega_{\rm ph}$, scales as $F_{\omega} (t \rightarrow \infty )\sim \beta t\sim N^2 t$, i.e., quadratically with respect to the emitted photons as clearly shown in Fig. \ref{fig:fig3} (b). Such a Heisenberg scaling outperforms significantly the one that can be achieved for sensing the trap frequency via the preparation of a coherent state with $N$ average excitations of the phonon mode, where the QFI scales as $I_{\rm coh} \sim N \sim{\sqrt{\beta}}$ corresponding to the SQL; cf  Fig. \ref{fig:fig3} (b). We emphasize that while such a Heisenberg scaling can be achieved by using the highly non-classical state $|\psi \rangle_{\rm nc}=\frac{1}{\sqrt{2}} (|0\rangle+|N\rangle)$ of the phonon mode, the preparation. of such state for large enough $N$ is however very challenging in practice.

Finally, in the proposed continuous-monitoring setting, the processing the continuously detected signal $D(t,0)$ in a single experimental run via Eq. (\ref{eq:sme}) allows us to construct the likelihood function $L_{\omega}[D(t,0)]=P_{\omega} [D(t,0)]/\int d \omega P_{\omega} [D(t,0)] $ which in turn provides us with an estimate of the unknown parameter $\omega$ via a simple maximum-likelihood strategy as illustrated in Fig. \ref{fig:fig5}.

\vspace{7mm}

\noindent \textbf{Discussion} 

\noindent Our study has proposed and analyzed the implementation of criticality-enhanced quantum sensing with single trapped ions via continuous measurement. A key innovation of our scheme is the use of a co-trapped ancillary ion as a detection module, allowing for highly efficient continuous monitoring (e.g., phonon counting) of the critical sensor. We have shown that the criticality enhanced precision scaling persists for imperfect detection efficiency, and is robust against various sources of noise in realistic experimental setups. Our proposal allows for a clear demonstration of criticality-enhanced precision \emph{scaling} well beyond the standard quantum limit, paving the way for harnessing driven-dissipative criticality to build next generation ultra-precise sensor devices, which may play a key role in electric-field imaging \cite{Bian:2021aa} and noise diagnosing near metallic surfaces \cite{RevModPhys.87.1419}.

\newpage

\noindent \textbf{METHODS} 

\noindent \textbf{Numerical calculation of the classical Fisher Information}

\noindent As explained in the Results, in our protocol the measured signal up to time $T=n dt$, $D(t,0)$, consists of a collection of the detected photons emitted from the strong transition $|d\rangle \rightarrow |e\rangle$ i.e $D(t,0) := \{ dN(n dt), \cdots, dN(dt),dN(0)\}$, where $dN(t)$ is a stochastic increment which for every time step $dt$ can take two values: $dN(t)=1$ with a modified probability $\tilde{p}_1=\epsilon \Gamma_{\rm s}  {\rm tr} \{ |d\rangle \langle d|\rho_{\rm c}\}dt$, where $\rho_{\rm c}\equiv\tilde{\rho}_{\rm c}/{\rm tr}(\tilde{\rho}_{\rm c})$ is the normalized conditional density matrix, while if there is no photon detection $dN(t)=0$ with probability $\tilde{p}_0=1-\tilde{p}_1$. Such a conditional evolution gives rise to the stochastic master equation  of the unnormalized-state written in Eq. (\ref{eq:sme}). The process of such signal gives rise to an estimator value of the bosonic mode frequency, $\omega$, whose precision in quantified by the classical Fisher Information (FI) as written in Eq. (\ref{eq:FI}):

\begin{equation}
\label{eq:FI2}
F_\omega(t) = \sum_{D(t,0)}P[D(t,0)]\left\{\partial_\omega {\rm{ln}} P[D(t,0)] \right\}^2.
\end{equation}
Crucially, the FI can be calculated numerically by approximating the ensemble average $\sum_{D(t,0)}P[D(t,0)]$ by a statistical average over sufficient (but finite) number of trajectories which are generated by typical Monte Carlo techniques following the numerical propagation of the normalized version of Eq.  (\ref{eq:sme}).  For each generated trajectory we can in turn calculate the respective probability $P[D(t,0)]={\rm tr \{\tilde{\rho}_{\rm c} \}}$ for slightly different values of $\omega$ up to an arbitrary multiplication factor, $C$, of our choice. Consequently, the $\left\{\partial_\omega {\rm{ln}} P[D(t,0)] \right\}^2$ can be extracted by a subsequent numerical differentiation (which is independent of $C$) and thus the FI can be estimated. We note that although straightforward, such approach  requires the propagation of a stochastic master equation  which can be numerically very demanding especially for large system sizes. 

As an alternative, we follow an equivalent strategy which gives rise to the same FI but requires only the propagation of a stochastic Schr{\"o}dinger  equation (SSE) and thus is numerically more efficient. In particular, we unravel the detected strong transition $|d\rangle \rightarrow |e\rangle$ with detection efficiency $\epsilon$ and rate $\Gamma_{\rm s}$, into two possible channels: i) one which is associated to the collapse of the conditional unnormalized state of the system into a``bright" channel $|\tilde{\Psi}_{\rm c} \rangle \rightarrow \sqrt{\epsilon \Gamma_{\rm s} dt} |e\rangle \langle d | \tilde{\Psi}_{\rm c} \rangle$ with probability $p_{\rm b}=\epsilon \Gamma_s dt \langle \Psi_c| d \rangle \langle d | \Psi_c \rangle$  where $| \Psi_c \rangle := | \tilde{\Psi}_c \rangle / \sqrt{\langle \tilde{\Psi}_{\rm c} | \tilde{\Psi}_{\rm c} \rangle}$ and ii) a second which is associated to the collapse of the conditional unnormalized state of the system into a ``dark" channel $|\tilde{\Psi}_{\rm c} \rangle \rightarrow \sqrt{(1-\epsilon)\Gamma_{\rm s} dt} |e\rangle \langle d | \tilde{\Psi}_c \rangle$ with probability $p_d=(1-\epsilon) \Gamma_s dt \langle \Psi_c| d \rangle \langle d | \Psi_c \rangle$. Following a quantum jump approach \cite{RevModPhys.70.101} we further unravel the undetected weak transition $|e\rangle \rightarrow |g\rangle$ into individual quantum trajectories where at each time step there is probability $p_{\rm un}= \Gamma_{\rm w} dt \langle \Psi_{\rm c}| e \rangle \langle e | \Psi_{\rm c} \rangle$ that the state of the system collapses into the ground state i.e $|\tilde{\Psi}_{\rm c} \rangle \rightarrow \sqrt{\Gamma_{\rm w} dt} |g\rangle \langle e | \tilde{\Psi}_{\rm c} \rangle$.  Consequently, at any infinitesimal time step $dt$ there is probability $p_{\rm b}$ that the systems collapses into the "bright" channel, probability $p_{\rm d}$ that it collapses into the "dark" channel, probability $p_{\rm un}$ that it collapses to the ground state and finally probability $p_0 ^\prime=1-p_{\rm b}-p_{\rm d}-p_{\rm un}$ to evolve according to the non-Hermitian evolution $ |\tilde{\Psi}_{\rm c}  \rangle \rightarrow \left( \hat{1}-idt\hat{H}_{\rm eff} \right) |\tilde{\Psi}_{\rm c} \rangle$ where $\hat{H}_{\rm eff}=\hat{H}_{\rm R}+\hat{H}_{\rm {a} }-i\Gamma_{\rm w} |e\rangle \langle e|-i \epsilon \Gamma_{\rm s} |d \rangle \langle d|-i(1-\epsilon) \Gamma_{\rm s} |d \rangle \langle d|$.  Quantitatively, the dynamics of the conditional unnormalized state, $|\tilde{\Psi}_{\rm c}\rangle$, can described in terms of a SSE
\begin{align}
\label{eq:SSE}
d | \tilde{\Psi}_{\rm c} \rangle&=- idt \hat{H}_{\rm eff} | \tilde{\Psi}_{\rm c} \rangle+dN_{\rm b}(t) \left(\sqrt{\epsilon \Gamma_{\rm s} dt}|e\rangle \langle d|-\hat{1}\right) | \tilde{\Psi} _{\rm c}\rangle \nonumber \\
&+ dN_{d}(t) \left(\sqrt{(1-\epsilon) \Gamma_{\rm s} dt}|e\rangle \langle d|-\hat{1}\right) | \tilde{\Psi} _{\rm c}\rangle \\ 
&+dN_{\rm un}(t) \left(\sqrt{\Gamma_{\rm w} dt } |g\rangle \langle e|-\hat{1}\right) | \tilde{\Psi} _{\rm c}\rangle \nonumber,
\end{align}
where $dN_j(t)$ with $j:=\{b,d, {\rm un} \}$is a stochastic increment associated with the respective collapse of the system and can take two values: $dN_j(t)=1$ with probability $p_j$ and $dN_j (t)=0$ with probability $1-p_j$. Notice that the probability of two counts in a time interval $dt$ vanishes faster than $dt$ and thus at any time interval we have at most one count in of the possible channels. In such approach the measured signal, $D^\prime (t,0)$, and thus the single trajectory generated via Monte Carlo methods is a collection of which ``jump" occurred at each time step i.e $D^\prime (t,0):= \{ \Big(dN_{\rm b}(ndt), dN_{\rm d}(ndt), dN_{\rm un}(ndt) \Big),\cdots $ $\Big(dN_{\rm b}(dt), dN_{\rm d}(dt), dN_{\rm un}(dt) \Big),\Big(dN_{\rm b}(0), dN_{\rm d}(0), dN_{\rm un}(0) \Big)\}$ and requires only the numerical propagation of the normalized version of the SSE in Eq. (\ref{eq:SSE}). Fixing the trajectory we can in turn neglect the ``jumps" happened in the ``dark" or the undetected channel, construct the trajectory $D^{\prime}_{\rm b} (t,0):= \{dN_{\rm b}(n dt) \cdots, dN_{\rm b}(0) \} \subseteq D^{\prime} (t,0)$ corresponding only to the collapse of the system in the  ``bright" channel and finally calculate numerically $\left\{\partial_\omega {\rm{ln}} P[D^{\prime}_b(t,0)] \right\}^2$. Therefore, we construct the corresponding FI 

\begin{equation}
\label{FI_sse}
F^{\prime}_\omega(t) = \sum_{D^{\prime}(t,0)}P[D^{\prime}(t,0)]\left\{\partial_\omega {\rm{ln}} P[D^{\prime}_{\rm b}(t,0)] \right\}^2,
\end{equation}
which is equal to Eq. (\ref{eq:FI2}), i.e $F^{\prime}_\omega(t)=F_\omega(t)$, for sufficiently enough numbers of generated trajectories.

\vspace{7mm}

\noindent \textbf{Quantifying the quality of the Scaling Collapse}

\noindent The finite-size scaling analysis is a powerful numerical method to extract the relevant exponents of continuous phase transitions. Consider a general quantity $\mathcal{A}$ dependent on two parameters $h$ and $L$ according to
\begin{equation}
\mathcal{A}(L,h)=h^{ a} f(h/L^{b}).
\label{eq:scaling}
\end{equation}
Depending on the nature of the model system, $\mathcal{A}$, $L$ and $h$ can refer to different quantities.  For example, $\mathcal{A}$ might refer to the magnetization of the Ising spin chain, with $h$ being the inverse of the transverse magnetic field and $L$  the length of the chain. In our case, $\mathcal{A}$ refers to the Fisher information $F_{\omega}(t, \beta)$, with $L$ and $h$ being the system size $\beta$ and the evolution time $t$ respectively.  From Eq.~(\ref{eq:scaling}), it is clear that if we plot $\mathcal{A} h^{-a}$ against $h/L^{b}$ for different values of $L$ and $h$, all the curves collapse onto a single. Therefore, we can determine the unknown exponents $a$ and $b$ by numerical fitting of the data to Eq.~\eqref{eq:scaling} and find the best scaling collapse. 

We can define appropriate measure of the quality of the data collapse to remove subjectiveness of the approach. If the scaling function $f(x)$ is known, we can define the measure~\cite{Somendra}
\begin{equation}
\label{eq:measure_data_collapse1}
\mathcal{M}_{\rm kn}(a,b)=\sqrt{\frac{1}{N} \sum_{ij} \left( \frac{\mathcal{A}({L_i,h_j})h_j^{-a}-f(L_i^{-b}h_j)}{f(L_i^{-b}h_j)}\right)^2}
\end{equation}
which is minimized by the optimal $a$ and $b$, with $N$ the total number of data points. Notice that in Eq.~(\ref{eq:measure_data_collapse1}) the division with respect to the scaling function $f(x)$ is essential---otherwise $\mathcal{M}_{\rm kn}$ can be minimized for small values of the numerator which not necessarily indicates a good collapse.

In the general case the scaling function $f(x)$ is not known. We can interpolate $f(x)$ via any set of data points corresponding to a specific $L$. Denoting the different sets via the subscript $p$, we can measure the quality of the data collapse by comparing the data points of pairs of sets $p_{1,2}$ in their overlapping regions and summing up all the contribution,
\begin{equation}
\label{eq:mesure_data_collapse2}
\mathcal{M}(a,b)=\sqrt{ \frac{1}{N}\sum_{p} \sum_{i\neq p} \sum_{j, \rm{ov}} \left(\frac{\mathcal{A}({L_i,h_j})h_j^{-a}-\mathcal{E}_p(L_i^{-b}h_j)}{\mathcal{E}_p(L_i^{-b}h_j)}\right)^2},
\end{equation}
where $\mathcal{E}_p(x)$ is the interpolation function based on the basis set $p$ and $N$ is the total number of points in all overlapping regions. Here, the innermost index $j$ runs over the overlapping region of the set $p$ and another set $i$. It is clear that   $\mathcal{M} \geq 0$ with the zero lower bound achieved only in case of perfect data collapse. As a result, it can be used via the variational principle for an automatic and objective extrapolation of the relevant critical exponents.

For our case, in order to examine the quality of the finite-size scaling analysis for different decoherence rates, as shown in Fig. 4(b) in the main text, we introduce the quality factor
\begin{equation}
Q=\frac{\mathcal{M}^F_{\rm id}} {\mathcal{M}^{\rm F}},
\end{equation}
where $ \displaystyle \mathcal{M}^{\rm F}=\mathcal{M}(2,1)=\sqrt{ \frac{1}{N}\sum_{p} \sum_{i\neq p} \sum_{j, \rm{ov}} \left(\frac{F (t_j \beta_i) t_j^2-\mathcal{E}_p(\beta_i t_j)}{\mathcal{E}_p(\beta_i t_j)}\right)^2}$ is the measure of the data collapse, and $\mathcal{M}^F_{\rm id}$ refers to the measure of the ideal noiseless case.

\vspace{7mm}

\noindent \textbf{Data availability}
The data sets generated and analyzed during the current study are available from the authors on reasonable request.

\vspace{7mm}

\noindent \textbf{Code availability}
All the numerical data presented in this paper are results of Python simulations and the code used to generate this data will be made available to the interested reader upon reasonable request.

\vspace{7mm}

\noindent \textbf{Acknowledgements} 

\noindent This work was supported by the EU project QuMicro (Grant No. 101046911) and the ERC Synergy grant HyperQ (grant no 856432). We acknowledge support by the state of Baden-W{\"u}rttemberg through bwHPC and the German Research Foundation (DFG) through grant no INST 40/575-1 FUGG (JUSTUS2 cluster). Part of the numerical simulations were performed using the QuTiP library~\cite{JOHANSSON20131234}.

\vspace{7mm}

\noindent \textbf{Competing interests} 
The authors declare no competing interests.

\vspace{7mm}

\noindent \textbf{Author contributions} 

\noindent D.Y., S.F.H. and M.B.P. designed the research. T.I. and D.Y. developed the model. T.I. performed the analytical calculations and numerical work with advice from  D.Y., S.F.H. and M.B.P. All authors discussed the results and wrote the manuscript. 

\newpage

\noindent \textbf{References} 

\bibliography{ContinuousMeasurementProposal_bib}
\onecolumngrid
\newpage
\section{Supplementary note 1: Realization of the phonon detector module }
\label{app:phonon_detector_module}
\begin{figure}[h]
\centering{} \includegraphics[width=0.85\textwidth]{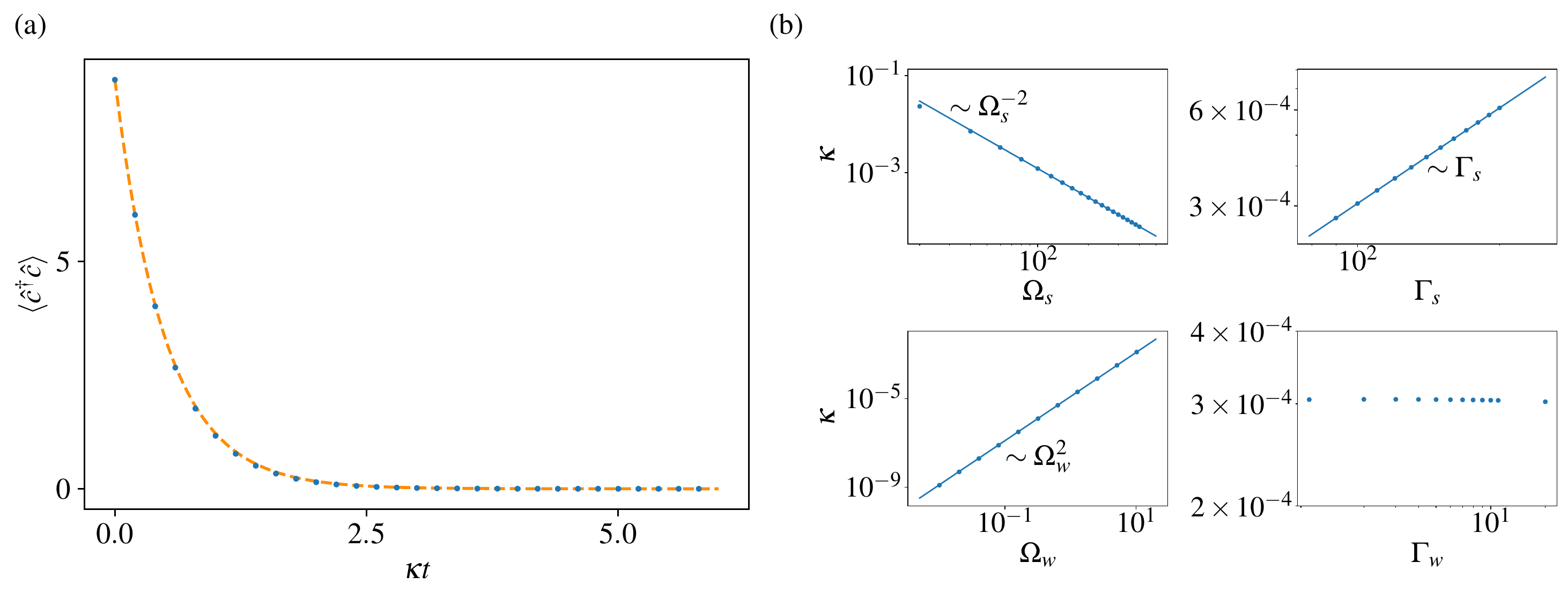} 
\caption{ Induced dissipation of the bosonic mode via a three level ancilla-ion. a) Comparison of the evolution of the phonon's excitation between an induced dissipation via an ancilla-ion with $\Gamma_{\rm w}=40$, $\Omega_{\rm w}=14.3 \Gamma_{\rm w}$, $\Gamma_{\rm s}=80 \Gamma_{\rm w}$, $\Omega_{\rm s}=2 \Gamma_{\rm s}$ and $\eta_{\rm LD}^{(2)}=0.07$ (blue dotted line) and a direct dissipation with $\kappa=1.5\times10^{-3} \Gamma_{\rm w}$ (orange dashed line).  b) Dependence of the induced dissipation rate, $\kappa$, with respect to different adjustable parameters in our system. The plots indicate that $\kappa \sim \Omega_{\rm w}^2 \Gamma_{\rm s}/\Omega_{\rm s}^2$ as expected from our physical understanding.}
\label{fig:supp_fig1} 
\end{figure}

In this section we discuss in detail the implementation of the phonon detector module with the help of the ancilla-ion, which has an internal ladder-type three-level structure as shown in Fig. 1 of the main text. Neglecting the system-ion, the Hamiltonian for the internal transition and the external motion (i.e., the phonon mode) of the ancilla reads ($\hbar=1$)
\begin{equation}
\label{eq:eq_ap_1}
\hat{H}_{\rm {a}}^{(I)}(t)=\omega_{\rm ph} \hat{c}^{\dagger}\hat{c}+\sum_{i=\rm g,e,d} \omega_i | i\rangle \langle i |+ \Omega_{\rm w} (|e\rangle \langle g|+\text{h.c.} ) \left[ e^{i (k_{\rm w} \hat{x}-\omega_{\rm w} t-\phi_{\rm w})} +\text{h.c.} \right]+\Omega_{\rm s} (|d\rangle \langle e|+\text{h.c.} ) \left[ e^{i (k_{\rm s} \hat{x}-\omega_{\rm s} t-\phi_{\rm s})} +\text{h.c.} \right],
\end{equation}
where the weak (strong) laser beam is characterised by the Rabi frequency $\Omega_{\rm w (s)}$, frequency $\omega_{\rm w (s)}$, wavevector $k_{\rm w (s)}$ and phase $\phi_{\rm w (s)}$. Here $\hat{x}=x_0 (\hat{c}+\hat{c}^{\dagger})$ is the ancilla-ion position operator, with $x_0=1/\sqrt{2 \omega_{\rm ph}m_{\rm{a} }}$ and $m_{\rm {a} }$ the mass of the ion. We assume that the Lamb-Dicke parameter of the two lasers $\eta_{w (s)} :=k_{w (s)} x_0$ are the same and denote them as $\eta_{\rm LD}^{(2)}$. Moving to a frame rotating with respect to $\hat{H}_0=\omega_{\rm ph} \hat{c}^{\dagger} \hat{c}+(\omega_e-\omega_g) |e \rangle \langle e|+(\omega_d-\omega_g) |d \rangle \langle d|$, the Hamiltonian Eq.~\eqref{eq:eq_ap_1} takes the form
\begin{align}
\hat{H}_{\rm {a}}(t) &= \Omega_{\rm w} (|e\rangle \langle g| e^{i(\omega_{\rm e}-\omega_{\rm g})t}+\text{h.c.} ) \left[ e^{i (\eta_{\rm LD}^{(2)} (\hat{c}e^{-i \omega_{\rm ph}t}+\hat{c}^{\dagger}e^{i \omega_{\rm ph}t})-\omega_{\rm w} t-\phi_{\rm w})} +\text{h.c.} \right] \\ 
&+\Omega_{\rm s} (|d\rangle \langle e| e^{i(\omega_d-\omega_e)t}+\text{h.c.} ) \left[ e^{i (\eta_{\rm LD}^{(2)} (\hat{c}e^{-i \omega_{\rm ph}t}+\hat{c}^{\dagger}e^{i \omega_{\rm ph}t})-\omega_{\rm s} t-\phi_{\rm s})} +\text{h.c.} \right]. \nonumber
\end{align}
In the Lamb-Dicke regime $ \eta_{\rm LD}^{(2)} \sqrt{\langle (\hat{c}+\hat{c}^{\dagger})^2 \rangle} \ll 1$ we have 
\begin{equation}
e^{\eta_{\rm LD}^{(2)} (\hat{c}e^{-i \omega_{\rm ph}t}+\hat{c}^{\dagger}e^{i \omega_{\rm ph}t})} \approx \mathcal{I}+ \eta_{\rm LD}^{(2)}(\hat{c}e^{-i \omega_{\rm ph}t}+\hat{c}^{\dagger}e^{i \omega_{\rm ph}t}).
\end{equation}
We can tune the two laser frequencies such that the weak laser drives resonantly the red phonon sideband of the $\ket{g}\to\ket{e}$ transition, $(\omega_{e\rm }-\omega_{\rm g})-\omega_{\rm w}=\omega_{\rm ph}$; while the strong laser drives resonantly the carrier transition $\ket{e}\to\ket{d}$, $(\omega_{\rm d}-\omega_{\rm e}) -\omega_{\rm s}=0$.  After choosing $\phi_{\rm s}=\phi_{\rm w}=0$, we perform an optical rotating-wave approximation (RWA) by neglecting terms rotating at frequency $\sim (\omega_{\rm e}-\omega_{\rm g})+\omega_{\rm w}$, $ \sim (\omega_{\rm d}-\omega_{\rm e})+\omega_{\rm s}$ and a vibrational RWA where we neglect terms rotating at frequency $\sim \omega_{\rm ph}$. As a result, we obtain
\begin{equation}
\label{eq:Hamiltonian_a-i}
\hat{H}_{\rm {a}}=\Omega_{\rm s} (| d\rangle \langle e|+|e\rangle \langle d |)+\eta_{\rm LD}^{(2)}\Omega_{\rm w}(|e\rangle \langle g| \hat{c}+|g\rangle \langle e| \hat{c}^{\dag}).
\end{equation}
for the Hamiltonian of the trapped ancilla-ion. 

The quantum state of the ancilla-ion $\rho_{\rm {a}}$ (including both the internal transition and  the external motion) evolves according to the LME
\begin{equation}
\label{eq:LME_a-i}
\frac{ d \rho_{\rm {a}} } {dt}= -i [\hat{H}_{\rm {a}},\rho_{\rm {a}}]+ \Gamma_{\rm s} \left( |e \rangle \langle d | \rho_{\rm {a}} |d \rangle \langle e| -\frac{1}{2} \{ |d\rangle \langle d|,  \rho_{\rm {a}}\}  \right)+\Gamma_{\rm w} \left( |g \rangle \langle e | \rho_{\rm {a}} |e \rangle \langle g| -\frac{1}{2} \{ |e\rangle \langle e |,  \rho_{\rm {a}} \}  \right).
\end{equation}
To visualize such an evolution, in Supplementary Figure~\ref{fig:supp_fig1}(a) we plot the phonon mode occupation in the appropriate parameter regime see below, which demonstrates an exponential decay with time at an effective dissipation rate $\kappa$. To acquire an understanding, let us relate the effective decay to the microscopic parameters $\Omega_{\rm s}$, $\Gamma_{\rm s}$, $\Gamma_{\rm w}$. We can divide one full cooling cycle $|g\rangle |n \rangle \rightarrow |g\rangle |n-1 \rangle$, where $|n \rangle$ denotes the phonon occupation number, into different elementary transitions. First, for small enough $\Gamma_{\rm w}$, the event $|g\rangle |n \rangle \rightarrow |e \rangle | n-1 \rangle $ defines the transition from the dark to the bright state of the ancilla ion at a characteristic timescale $T_1 \approx \Omega_{\rm s}^2/(\Omega_{\rm w}^2 \Gamma_{\rm s})$~\cite{RevModPhys.70.101}.  The inverse, undesired transition $|e\rangle |n-1 \rangle \rightarrow |g \rangle | n \rangle $, with a characteristic timescale $T_2 \approx (\Gamma_{\rm s}^2+\Omega_{\rm s}^2) T_1/ \Gamma_{\rm s}^2$, restores the initial state of the ancilla-ion but without any phonon annihilation. As a result, to realize an effective dissipation of the phonon mode, we require that the spontaneous emission $|e\rangle |n-1 \rangle \rightarrow |g\rangle |n-1 \rangle$ happens at a timescale $ 1/\Gamma_{\rm w} \equiv T_3  \ll T_2$.  The time of a complete cooling cycle is given by $1/\kappa \equiv T_c\approx T_1+T_3 \approx T_1$ as we are in the regime where $\Omega_{\rm s} \gg \Omega_{\rm w}$. Such analytical understanding is confirmed by the numerical simulation shown in Supplementary Figure~\ref{fig:supp_fig1}(b), where we plot the dependence of  the effective dissipation rate $\kappa$ of the phonons with respect to the different parameters of our model.

\section{Supplementary Note 2: Trapped-ion implementation of the open Rabi model }
\label{app:real_rabi_model}

\begin{figure}[h]
\centering{} \includegraphics[width=0.6\textwidth]{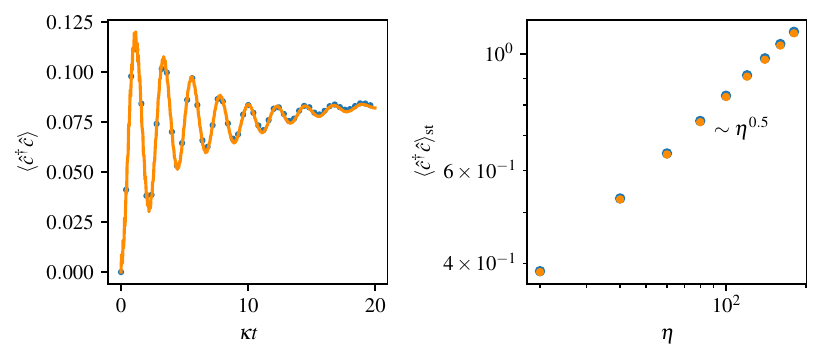} 
\caption{ Trapped-ion realisation of the open Rabi model. a) Dynamics of the phonon's excitation in the open Rabi model as implemented with the help of an ancilla-ion and evolved according to Eq. (\ref{eq:LME_a-i_s-i}) (blue-dotted line) and Eq. (2) of the main text (orange solid line).  Parameters: $\beta=50$, $\omega=0.6$; the rest kept the same as Supplementary Figure~\ref{fig:supp_fig1} .  b) Steady state occupation of the bosonic mode with respect to different system sizes, $\beta$, as calculated from Eq.(\ref{eq:LME_a-i_s-i})  (blue-dotted line) and Eq. (2) of the main text (orangle-dotted line). The excellent agreement between the two lines in both plots demonstrates the accurate implementation of the open Rabi model in our set-up.}
\label{fig:supp_fig2} 
\end{figure}

Here, we discuss the implementation of the open Rabi model as presented schematically in Fig. 1 of the main text. The system-ion is co-trapped with the ancilla-ion analyzed in Appendix \ref{app:phonon_detector_module}, and is driven by two travelling-wave laser beams of the same Rabi frequency, $\Omega_0$, and Lamb-Dicke parameter, $\beta_{\rm LD}$. As a result the Hamiltonian of the total system reads

\begin{equation}
\hat{H}_{\rm tot}^{(I)}(t)=\hat{H}_{\rm {a}}^{(I)}(t)+\hat{H}_{\rm {s}}^{(I)}(t),
\end{equation}
where $\hat{H}_{\rm {s}}^{(I)}(t)=\frac{\omega_I}{2}\hat{\sigma}_{\rm z}+\sum_j \frac{\Omega_0}{2} \hat{\sigma}_{\rm x} \left( e^{i \eta_{\rm LD} (\hat{c}+\hat{c}^{\dagger})-\omega_j t -\phi_j }+\text{h.c.} \right)$ and $\hat{\sigma}_{\rm z,x}$ are the Pauli matrices referring to the system-ion. In the rotating frame with respect to $\tilde{H}_0=\omega_{\rm ph} \hat{c}^{\dagger} \hat{c}+(\omega_{\rm e}-\omega_{\rm g}) |e \rangle \langle e|+(\omega_{\rm d}-\omega_{\rm g}) |d \rangle \langle d|+\frac{\omega_I}{2}\hat{\sigma}_{\rm z}$, following similar approximations as before by choosing $\omega_1=\omega_{\rm I}+\omega_{\rm ph}-\delta_{\rm b}$, $\omega_2=\omega_{\rm I}-\omega_{\rm ph}-\delta_{\rm r}$,    $\phi_1=\phi_2=3 \pi /2$, $\omega_{\rm s}=(\omega_{\rm d}-\omega_{\rm g})$, $(\omega_{\rm e}-\omega_{\rm g})-\omega_{\rm w}=\omega_{\rm ph}+(\delta_{\rm b}-\delta_{\rm r})/2$ we arrive at 
\begin{align}
\hat{H}_{\rm tot}^{(I)}(t)&=\Omega_{\rm s} (| d\rangle \langle e|+|e\rangle \langle d |)+\eta_{\rm LD}^{(2)}\Omega_{\rm w}(|e\rangle \langle g| \hat{c}e^{i\frac{\delta_{ \rm b}-\delta_{\rm r}}{2}t}+|g\rangle \langle e| \hat{c}^{\dag}e^{-i\frac{\delta_{ \rm b}-\delta_{\rm r}}{2}t})\nonumber \\
&-\frac{\Omega_0 \eta_{\rm LD}}{2} ( \hat{\sigma}_+ e^{i\frac{\delta_{ \rm b}+ \delta_{ \rm r} }{2}t}+\text{h.c})(\hat{c}^{\dagger} e^{i\frac{\delta_{ \rm b}- \delta_{ \rm r} }{2}t}+\text{h.c}).
\label{eq:eq_ap_7}
\end{align}
In the frame rotating with $H^{'}_0=\frac{\delta_{\rm b}-\delta_{\rm r}}{2}\hat{c}^\dag\hat{c}+\frac{\delta_{\rm r}+\delta_{\rm b}}{4}\hat{\sigma}_z$, the evolution of the full model is governed by the LME
\begin{equation}
\label{eq:LME_a-i_s-i}
\frac{ d \rho } {dt}= -i [\hat{H}_{\rm {s}}+\hat{H}_{\rm {a}},]+ \Gamma_{\rm s} \left( |e \rangle \langle d | \rho |d \rangle \langle e| -\frac{1}{2} \{ |d\rangle \langle d|,  \rho\}  \right)+\Gamma_{\rm w} \left( |g \rangle \langle e | \rho |e \rangle \langle g| -\frac{1}{2} \{ |e\rangle \langle e |,  \rho \}  \right),
\end{equation}
which correctly implements the LME (Eq. (2) of the main text) of the open Rabi model as illustrated in Supplementary Figure~\ref{fig:supp_fig2}. Although Eq. (\ref{eq:LME_a-i_s-i}) describes the dynamics of the unconditional state of the ancilla and the system ion, following a standard procedure \cite{gardiner2} one arrives to the conditional stochastic master equation (SME) written in Eq. (8) of the main text, given that we do not detect the weak transition $|e\rangle \rightarrow |g\rangle$ and we detect the strong transition $|d\rangle \rightarrow |e\rangle$ with efficiency $\epsilon$.

\section{Supplementary Note 3: Enhanced Continuous Phonon Detection}
\label{app: en_con_ph_det}
In this section we discuss the highly efficient continuous measurement of the phonon mode accomplished by the use of the ancilla-ion. Adopting the physical picture developed in Appendix \ref{app:real_rabi_model}, it is clear that once an event $|g\rangle |n \rangle \rightarrow |e \rangle | n-1 \rangle $ occurs, the strongly driven transition $|e\rangle \leftrightarrow | d\rangle$ of the ancilla-ion leads to the emission of photons at a rate $\Gamma_{\rm ph}=\Gamma_{\rm s} \times P_{\rm ex} \approx  \frac{\Gamma_{\rm s}}{2} \frac{(2 \Omega_{\rm s}/\Gamma_{\rm s})^2}{1+(2 \Omega_{\rm s}/\Gamma_{\rm s})^2}$, where $P_{\rm ex}$ is the probability to find the ancilla-ion in the $|d\rangle \langle d|$ state. The characteristic timescale for restoring the population, $|e\rangle \rightarrow |g\rangle$, is $T_2=1/\Gamma_{\rm w}$. As a result, during a full cooling cycle $|g\rangle |n\rangle \rightarrow |g\rangle |n-1\rangle$ there are $N_{\rm em}=T_2 \times \Gamma_{\rm ph}\approx \Gamma_{\rm s}/\Gamma_{\rm w}$ photons emitted and $N_{\rm ph}=\epsilon N_{\rm em} $ photons detected which lead to the definition of the enhancement factor as in Eq. (7) of the main text. As a simple demonstration, let us again neglect the system-ion, and consider the enhanced measurement of a single vibrational phonon mode occupying an excited Fock state. The conditional evolution of the phonon mode is given by a stochastic master equation (SME) similarly to Eq. (8) of the main text. As shown in Supplementary Figure~\ref{fig:supp_fig3}, for any photon detector of efficiency, $\epsilon$, we can implement almost perfect detection of the phonon mode by appropriate engineering of the parameters for the internal transitions the ancilla.

\begin{figure}[h]
\centering{} \includegraphics[width=0.5\textwidth]{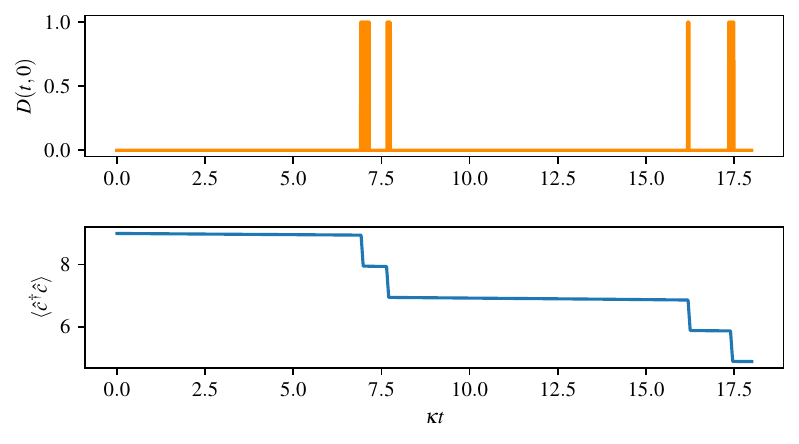} 
\caption{Conditional evolution of the damped bosonic mode of the trap as implemented with the help of an ancilla-ion where each annihilated phonon is accompanied with the emission and detection of lots of photons. Upper panel: The photon-detection signal, $D(t,0)$, of a detector with efficiency $\epsilon=0.5$; Lower panel: the conditional evolution of the damped mode as exemplified by the bosonic mode occupation $\langle \hat{c}^{\dagger} \hat{c} \rangle$. The dynamics of a single trajectory associated with the corresponding detected signal is the same as the ideal case of perfect photon detection. Parameters: $\Gamma_{\rm w}=40$, $\Omega_{\rm w}=14.3\Gamma_{\rm w}$, $\Omega_{\rm s}=2\Gamma_{\rm s}=160 \Gamma_w$ and $\eta_{\rm LD}^{(2)}=0.07$.}
\label{fig:supp_fig3} 
\end{figure}

\end{document}